\documentclass[reprint,amsmath,amssymb,jcp,aip,floatfix]{revtex4-2}
\usepackage[utf8]{inputenc}
\usepackage[T1]{fontenc}
\usepackage{mathptmx}
\usepackage{etoolbox}
\usepackage{graphicx}
\usepackage{dcolumn}
\usepackage{bm}
\usepackage[version=4]{mhchem}
\usepackage[group-separator={,}]{siunitx}
\usepackage{blkarray}
\usepackage{braket}
\usepackage[hidelinks]{hyperref}
\DeclareSIUnit\angstrom{\text {Å}}
\DeclareSIUnit\atomicmassunit{u}
\DeclareSIUnit\elementarycharge{\text {\ensuremath {e}}}
\DeclareSIUnit\bohr{\text {\ensuremath {a}}_{0}}

\newcommand{\dd}{\mathrm{d}}

\makeatletter
\def\@email#1#2{%
 \endgroup
 \patchcmd{\titleblock@produce}
  {\frontmatter@RRAPformat}
  {\frontmatter@RRAPformat{\produce@RRAP{*#1\href{mailto:#2}{#2}}}\frontmatter@RRAPformat}
  {}{}
}%
\makeatother
\begin{document}

\title[Scalar machine learning of tensorial quantities -- Born effective charges from monopole models]{Scalar machine learning of tensorial quantities -- Born effective charges from monopole models}

\author{Bernhard Schmiedmayer}
\email{bernhard.schmiedmayer@univie.ac.at}
\affiliation{%
University of Vienna, Faculty of Physics and Center for Computational Materials Science, Kolingasse 14-16, A-1090 Vienna, Austria
}

\author{Angela Rittsteuer}
\affiliation{%
University of Vienna, Faculty of Physics and Center for Computational Materials Science, Kolingasse 14-16, A-1090 Vienna, Austria
}
\affiliation{%
Vienna Doctoral School in Physics, Kolingasse 14-16, A-1090 Vienna, Austria
}

\author{Tobias Hilpert}
\affiliation{%
University of Vienna, Faculty of Physics and Center for Computational Materials Science, Kolingasse 14-16, A-1090 Vienna, Austria
}
\affiliation{%
Vienna Doctoral School in Physics, Kolingasse 14-16, A-1090 Vienna, Austria
}

\author{Georg Kresse}
\affiliation{%
University of Vienna, Faculty of Physics and Center for Computational Materials Science, Kolingasse 14-16, A-1090 Vienna, Austria
}
\affiliation{%
VASP Software GmbH, Berggasse 21/14, A-1090, Vienna, Austria
}

\date{\today}

\begin{abstract}
   Predicting tensorial properties with machine learning models typically requires carefully designed tensorial descriptors. In this work, we introduce an alternative strategy for learning tensorial quantities based on scalar descriptors. We apply this approach to the Born effective charge tensor, showing that scalar (monopole) kernel models can successfully capture its tensorial nature by exploiting the definition of the Born effective charge tensor as the derivative of the polarisation with respect to atomic displacements. We compare this method with tensorial (dipole) kernel models, as established in our previous work, in which the tensorial structure of the Born effective charge is encoded directly in the kernel and obtained via its derivative. Both approaches are then used for charge partitioning, enabling the separation of monopole and dipole contributions. Finally, we demonstrate the effectiveness of the framework by computing finite-temperature infrared spectra for a range of complex materials.
\end{abstract}

\maketitle
\section{Introduction}
\label{sec:intro}
Recent advances in machine learning have paved the way for new developments in the field of computational materials science. Among the most significant advances is the construction of high-fidelity, transferable machine-learned force fields (MLFFs), which can generate long-timescale molecular dynamics trajectories.\cite{bartok2010gaussian,morawietz2016van,bonati2018silicon,behler2007generalized,bartok2018machine,jinnouchi2019fly,reiser2022graph,wu2020comprehensive,gasteiger2020directional,haghighatlari2022newtonnet,takamoto2022teanet,batzner20223,batatia2022mace,corso2024graph} These developments allow structural and dynamical properties to be predicted under realistic conditions, thereby bridging the gap between simulations and experiments.\cite{gastegger2017machine,jinnouchi2019phase,sommers2020raman,zhang2020efficient,verdi2021thermal,jansen2023phase,schienbein2023spectroscopy,montero2024comparing,schmiedmayer2024derivative,falletta2025unified,schmiedmayer2025equivariant} A significant challenge in this context is how to treat tensorial quantities, such as atomic polarisation or the Born effective charge (BEC), which must obey transformation rules in order to remain physically meaningful and independent of the choice of the reference frame.\cite{schutz1980geometrical,nakahara2018geometry}

Equivariant architectures address this issue by incorporating symmetry constraints directly into the model. This ensures that features and outputs are correctly transformed under rotations and translations in three-dimensional Euclidean space.\cite{worrall2017harmonic,grisafi2018symmetry,geiger2022e3nn} These approaches have already improved the accuracy of interatomic potentials\cite{batzner20223,batatia2025foundation} and enabled the prediction of optical, phonon, and scattering spectra.\cite{schuett2021equivariant,cheng2023direct,okabe2024virtual} Furthermore, they have enabled the prediction of tensorial materials properties, including elasticity and electronic response.\cite{venetos2023machine,pakornchote2023straintensornet,wen2024equivariant,schmiedmayer2024derivative,schmiedmayer2025equivariant} Recently, equivariant machine learning frameworks have been used to predict the BEC tensor. These frameworks include symmetry-restricted neural networks,\cite{shimizu2023prediction} equivariant graph convolutional neural networks,\cite{kutana2025representing} and derivative learning with deep potentials,\cite{zhang2022deep} kernel-based regression,\cite{schmiedmayer2024derivative} as well as differentiable deep neural networks,\cite{malenfant2024efficient} and graph convolutional neural networks.\cite{falletta2025unified} Recent studies by the group of B.~Cheng have employed a related strategy to model BECs using monopole-based representations.\cite{kim2025universal,zhong2025machine} In these studies, the learning of BECs is incorporated directly into a long-range machine-learning force field through a Latent Ewald Summation framework.\cite{cheng2025latent} The learned charges are used for the evaluation of long-range electrostatic energies, polarisation, and their derivatives. In addition, a number of other approaches have been proposed that model BECs within equivariant machine-learning frameworks.\cite{sommers2020raman,zhang2020efficient,gastegger2017machine,schienbein2023spectroscopy,joll2024molecular}

Building on our previous derivative-learning strategy for predicting finite-temperature infrared spectra from BECs (see Ref.~\onlinecite{schmiedmayer2024derivative}), we demonstrate that the same physical constraints can be enforced using purely scalar machine-learning models by expressing the BEC tensor through a monopole--dipole response decomposition. The approach relies exclusively on invariant descriptors and models, specifically SOAP\cite{bartok2013representing} and MACE.\cite{batatia2022mace} Tensorial equivariance is recovered implicitly by exploiting the definition of the BEC tensor as the derivative of the polarisation with respect to atomic displacements, together with a first-order multipolar expansion. While our previous work proved that derivative learning effectively solves the issue of arbitrary polarisation phase in bulk systems, it required equivariant kernels to learn the dipole vector directly. Here, we ask if simpler, invariant scalar descriptors can achieve similar accuracy by learning the monopole charge scalar instead.

\section{Method}
\label{sec:meth}
\subsection{General Remarks}
Previously, we addressed the indeterminacy of bulk polarisation by learning the BEC tensor as the derivative of a polarisation vector. Here, we extend this by decomposing the BEC tensor into a local rigid-ion term (scalar) and a charge-redistribution term (derivative of scalar), avoiding the need for equivariant machine-learning entirely. As already mentioned in the introduction, this is largely inspired by the derivation of B. Cheng that BECs are relatively reliably predicted using a latent representation of the interacting point charges. Here we concentrate on direct predictions of BECs without considering total energies. 

Within Kohn--Sham density functional theory, the total energy of a material or molecule with a set of atomic positions $\{\mathbf{r}_i\}$ in the presence of a smooth external electric potential $\phi(\mathbf{x})$ at Cartesian position $\mathbf{x}=\left(x^\alpha\right)$ an be exactly decomposed into the zero-field Kohn--Sham energy, $E_\text{KS}$, and the interaction energy, $E_\text{int}$, describing the coupling to the external potential,\cite{umari2002ab}
\begin{equation}
    E_\text{tot}(\{\mathbf{r}_i\},\phi) = E_\text{KS}(\{\mathbf{r}_i\})+E_\text{int}(\{\mathbf{r}_i\},\phi).
\end{equation}
The interaction energy between the potential and the system’s charge density $\rho(\{\mathbf{r}_i\},\mathbf{x})$ is
\begin{equation}\label{eq:EintS}
    E_\text{int}(\{\mathbf{r}_i\},\phi) = \int_{\mathbb{R}^3} \rho(\{\mathbf{r}_i\},\mathbf{x})\phi(\mathbf{x})\dd x^3.
\end{equation}
The total charge $Q$ of the system is given by the integral of the charge density,
\begin{equation}
    Q(\{\mathbf{r}_i\}) = \int_{\mathbb{R}^3} \rho(\{\mathbf{r}_i\},\mathbf{x})\dd x^3.
\end{equation}
Under a constant shift in the external potential, $\phi\to\phi+c$, the interaction energy transforms as $E_\text{int}\to E_\text{int}+cQ$. For a transferable and well-defined machine-learning model, this dependence on the arbitrary zero of the electrostatic potential is undesirable, at least for solids, as physical observables and response properties should depend only on the external electric field, $\mathcal{E}(\mathbf{x}) = -\nabla\phi(\mathbf{x})$, which is invariant under such a shift. Therefore, to ensure this gauge invariance, we will consider only systems with charge neutrality, {\em i.e.}, $Q(\{\mathbf{r}_i\})=0$.

We represent the charge density by a set of localised monopoles at positions $\{\mathbf{r}_k\}$, with corresponding charges $\{q_k\}$,
\begin{equation}
    \rho_\text{mon}(\{\mathbf{r}_i\},\mathbf{x}) = \sum_k q_k\left(\{\mathbf{r}_i\}\right) \delta(\mathbf{x}-\mathbf{r}_k).
\end{equation}
Here we explicitly allow the charges $q_k$ to depend on the atomic configuration $\{\mathbf{r}_i\}$. This reflects the physical fact that the electronic charges redistribute in response to atomic displacements. Inserting this form into the interaction energy yields
\begin{equation}\label{eq:Eint}
    E_\text{int}^\text{mon}\left(\{\mathbf{r}_i\},\phi\right) = \sum_k q_k\left(\{\mathbf{r}_i\}\right)\phi(\mathbf{r}_k).
\end{equation}

Now, we consider the first-order response of the interaction energy to a small displacement $\delta \mathbf{r}_j$ of atom $j$. This is given by the partial differential of Eq.~\eqref{eq:Eint},
\begin{equation}
    \delta E_\text{int}^\text{mon} = \frac{\partial E_\text{int}^\text{mon}}{\partial \mathbf{r}_j} \cdot \delta \mathbf{r}_j = \sum_k \left[ \frac{\partial q_k}{\partial \mathbf{r}_j} \phi(\mathbf{r}_k) + q_k \frac{\partial \phi(\mathbf{r}_k)}{\partial \mathbf{r}_j} \right] \cdot \delta \mathbf{r}_j.
\end{equation}
Recognising that $\partial\phi(\mathbf{r}_k)/\partial \mathbf{r}_j=-\mathcal{E}(\mathbf{r}_j) $ and using the definition of the external electric field, this simplifies to,
\begin{equation}
    \delta E_\text{int}^\text{mon} = \sum_k \phi(\mathbf{r}_k) \underbrace{\frac{\partial q_k}{\partial \mathbf{r}_j} \cdot \delta\mathbf{r}_j}_{\delta q_k} - q_j \mathcal{E}(\mathbf{r}_j) \cdot \delta \mathbf{r}_j.
\end{equation}
Here $\delta q_k$ is the response of the charge $q_k$ under small displacements $\delta \mathbf{r}_j$.   

To connect this response to macroscopic dielectric properties, we expand the external potential around the origin to first (dipole) order, which is exact for a uniform field $\mathcal{E}(0)$. This gives $\phi(\mathbf{r}_k) \approx \phi(0) - \mathbf{r}_k \cdot \mathcal{E}(0)$, which corresponds to retaining only the linear term in the Taylor expansion of the external potential. Consistently, the electric field is then spatially uniform, $\mathcal{E}(\mathbf{r}_j)\approx\mathcal{E}(0)$.\cite{jackson2013klassische} Substituting these into the energy response yields,
\begin{equation}\label{eq:deltaEint}
    \delta E_\text{int}^\text{mon} \approx \phi(0) \underbrace{\sum_k\delta q_k}_{\delta Q} - \mathcal{E}(0)\cdot \underbrace{\left[q_j\delta \mathbf{r}_j + \sum_k \mathbf{r}_k \delta q_k\right]}_{\delta \mathbf{P}_\text{mon}}.
\end{equation}
The first term contains $\sum_k\partial q_k/\partial\mathbf{r}_j=\partial Q / \partial\mathbf{r}_j$. That is the change in total charge $\delta Q$. For an isolated system, charge is conserved, so this term is zero. The second term contains $\delta \mathbf{P}_\text{mon}$, which is the change in the system's electric dipole moment $\mathbf{P}_\text{mon}(\{\mathbf{r}_i\})=\sum_k q_k \mathbf{r}_k.$ The change in interaction energy thus takes the familiar dipolar form,
\begin{equation}
    \delta E_\text{int}^\text{mon}(\{\mathbf{r}_i\},\phi) \approx - \mathcal{E}(0)\cdot\delta \mathbf{P}_\text{mon}(\{\mathbf{r}_i\}).
\end{equation}
For completeness, we recall that $\mathbf{P}$ is origin-dependent whenever the system carries a net charge $Q\neq 0$, since $\mathbf{P}_\text{mon}\left(\{\mathbf{r}_i+\mathbf{r}\}\right) = \mathbf{P}_\text{mon}\left(\{\mathbf{r}_i\}\right)+Q\mathbf{r}$. In the present case, we have already assumed charge neutrality ($Q=0$), so the polarisation is origin-independent.

Following the standard definition, the BEC tensor, $\mathbf{Z}^*_j$ of atom $j$ relates the macroscopic polarisation $\mathbf{P}$ to atomic displacements $\mathbf{r}_j$;\cite{morcillo1966infra,morcillo1969ir,ghosez1997dynamical,ghosez1998dynamical}
\begin{equation}
    Z_{j,\alpha\beta}^* = \frac{\partial P^\alpha}{\partial r^\beta_j} = -\frac{\partial^2 E_\text{tot}}{\partial \mathcal{E}^\alpha\partial r^\beta_j} \Bigg|_{\mathcal{E}=0}.
\end{equation}
An infinitesimal displacement induces a linear polarisation response,
\begin{equation}\label{eq:dieResp}
    \delta P^\alpha= \sum_\beta Z_{j, \alpha\beta}^* \delta r^\beta_j.
\end{equation}
From the derivation for monopoles above, we identify the explicit form of the tensor as;
\begin{equation}\label{eq:Zmol}
    Z_{j,\alpha\beta}^{*,\text{mon}} = q_j \delta_{\alpha\beta} + \sum_k r_k^\alpha\frac{\partial q_k}{\partial r_j^\beta}.
\end{equation}
This expression decomposes the dielectric response into two physically distinct contributions. The first term, $q_j \delta_{\alpha\beta}$, is the local, rigid-ion contribution, arising from the polarisation created when a static point charge $q_j$ is displaced. The second term, $\sum_k r_k^\alpha\partial q_k/\partial r_j^\beta$, is the non-local, charge-redistribution contribution. It accounts for the change in the dipole moment caused by the dynamic rearrangement of charge on all atoms in response to the displacement of just atom $j$. In first principles calculations, atomic displacements necessarily polarise the electronic density, making this charge-redistribution term crucial. If the charges $\{q_i\}$ were assumed to be fixed (the rigid-ion approximation), this second term would vanish, and the BECs would incorrectly reduce to the nominal ionic charges ($Z^*_j\to q_j$). The BEC in many mixed ionic--covalent crystals are found to be substantially different from their formal oxidation states.\cite{ghosez1998dynamical} The dynamic charge-redistribution term is precisely what accounts for the anisotropic behaviour, thus providing a more realistic and physically complete description of the polarisation response in materials.

For systems with periodic boundary conditions, such as liquids and solids, using absolute atomic positions is ill-defined. The polarisation and its derivatives must instead be formulated in terms of relative coordinates. Choosing the coordinate origin at $\mathbf{r}_i$, the monopole contribution to the BEC tensor of atom $i$, takes the form,
\begin{equation}\label{eq:Zsol}
    Z_{i,\alpha\beta}^{*,\text{mon}}(r^\alpha_i) = q_i\delta^{\alpha\beta} + \sum_{j \neq i} \frac{\dd q_j}{\dd r^\beta_i}(r_j^\alpha-r_i^\alpha).
\end{equation}
This expression highlights a limitation compared to our previous work. In the dipole-based approach, the BEC is the derivative of a single global polarisation $\mathbf{P}$. This ensures that $\mathbf{P}$ can be recovered as the anti-derivative of $\mathbf{Z}^*$. However, because Eq.~\eqref{eq:Zsol} constructs the BEC using charges $q_i$ computed from different centres, the resulting tensor $\mathbf{Z}^*$ does not possess a global anti-derivative. With this form, the system polarisation cannot be obtained by integrating $\mathbf{Z}^*$. For Eq.~\eqref{eq:Zsol} to be unique and independent of the choice of origin, two conditions must hold for the system as a whole,
\begin{equation}\label{eq:constraint}
    \sum_i q_i = 0
    \quad \text{and} \quad
    \sum_i \frac{\dd q_i}{\dd r^\beta_j} = 0
    \quad \forall i, \beta.
\end{equation}
The first condition enforces overall charge neutrality, while the second enforces that the total charge is conserved during any atomic displacement. With a limited set of traning data, that all individually observe $\sum_j \mathbf{Z}^*_j=0,$ we found that the charge neutrality in Eq.~\eqref{eq:constraint} is not automatically observed by the trained model. We hence found it expedient to constrain charge neutrality explicitly.

The model can be refined by including higher-order electrostatic interactions. To remain consistent with an external potential truncated at dipole order, we expand the charge density to include point-dipoles in addition to monopoles. A physical dipole can be represented as two monopoles of opposite charge, $\pm q_k$, separated by an infinitesimal displacement vector $\mathbf{d}$.\cite{fliessbach2012elektrodynamik} The corresponding charge density is
\begin{equation}
\begin{split}
    \rho_\text{dip}(\{\mathbf{r}_i\},\mathbf{x}) = \sum_k \Bigg[ q_k\left(\{\mathbf{r}_i\}\right) \delta\left(\mathbf{x}-\left(\mathbf{r}_k+\frac{\mathbf{d}}{2}\right)\right) \\
    -q_k\left(\{\mathbf{r}_i\}\right) \delta\left(\mathbf{x}-\left(\mathbf{r}_k-\frac{\mathbf{d}}{2}\right)\right) \Bigg].
\end{split}
\end{equation}
In the pure point-dipole limit, the separation $\mathbf{d}$ is an infinitesimal while the dipole moment $\mathbf{p}_k = q_k \mathbf{d}$ remains finite. Consequently, the charge magnitude $q_k$ must diverge. This is precisely the definition of a directional derivative, giving the expression for the point dipole density
\begin{equation}
    \rho_\text{dip}(\{\mathbf{r}_i\},\mathbf{x}) = \sum_k \underbrace{q_k\mathbf{d}}_{\mathbf{p}_k}\cdot\nabla\delta(\mathbf{x}-\mathbf{r}_k).
\end{equation}
Using this density in Eq.~\eqref{eq:EintS}, the dipolar part of the interaction energy is
\begin{equation}
    E_\text{int}^\text{dip}\left(\{\mathbf{r}_i\},\phi\right) = \sum_k \mathbf{p}_k\left(\{\mathbf{r}_i\}\right)\cdot\mathcal{E}(\mathbf{r}_k).
\end{equation}
A minus sign arises during integration by parts, but it is exactly absorbed when rewriting the result in terms of the electric field. The response of this dipolar interaction energy to an atomic displacement $\delta \mathbf{r}_j$ is
\begin{equation}
\begin{split}
    \delta E_\text{int}^\text{dip}\left(\{\mathbf{r}_i\},\phi\right) = \mathbf{p}_j\left(\{\mathbf{r}_i\}\right) \cdot \nabla\mathcal{E}(\mathbf{r}_j) \cdot \delta \mathbf{r}_j \\
    + \sum_k \frac{\partial \mathbf{p}_k\left(\{\mathbf{r}_i\}\right)}{\partial \mathbf{r}_j} \cdot \mathcal{E}(\mathbf{r}_k) \cdot \delta \mathbf{r}_j.
\end{split}
\end{equation}
Restricting to the dipole-order approximation, we evaluate the field at the origin --- as done above --- yielding
\begin{equation}
    \delta E_\text{int}^\text{dip}\left(\{\mathbf{r}_i\},\phi\right) \approx\mathcal{E}(0)\cdot\underbrace{\left[\sum_k \frac{\partial \mathbf{p}_k\left(\{\mathbf{r}_i\}\right)}{\partial \mathbf{r}_j} \cdot \delta \mathbf{r}_j\right]}_{\partial\mathbf{P}_\text{dip}}.
\end{equation}
From this, follows Eq.~\eqref{eq:dieResp} to identify the dipolar contribution to the BECs as
\begin{equation}\label{eq:Zdie}
    Z_{j,\alpha\beta}^{*,\text{dip}} = \sum_k \frac{\partial p^\alpha_k}{\partial r_j^\beta}.
\end{equation}
Finally, the total BEC tensor, including monopolar and dipolar contributions, is
\begin{equation}\label{eq:Z}
    Z_{i,\alpha\beta}^{*} = q_i\delta^{\alpha\beta} + \sum_{j\neq i} \frac{\dd q_j}{\dd r^\beta_i}(r_j^\alpha-r_i^\alpha) + \sum_k \frac{\partial p^\alpha_k}{\partial r_i^\beta}.
\end{equation}
Whereas this expression has no well-defined anti-derivative of $\mathbf{Z}^*_i$. However, Eq.~\eqref{eq:Zmol} can be extended to dipolar order. In this case, the polarisation $\mathbf{P}$ of the system is well-defined and can be written as
\begin{equation}
    P^\alpha = \sum_j \left[q_j r_j^\alpha+p_j^\alpha\right].
\end{equation}

\subsection{Kernel method}

In our kernel approach, we use a local-environment dependent kernel $K$ to model each atomic charge as\cite{bishop2006pattern}
\begin{equation}
q_i = \sum_{I_B} \omega_{I_B} K(X_i, X_{I_B}).
\end{equation}
Here $X_i$ denotes the feature vector representing the local environment of atom $i$, $\{X_{I_B}\}$ is a set of reference environments (sparse basis functions in which $K$ is represented), and $\{\omega_{I_B}\}$ are the fitting coefficients of the machine-learning model. The simplest example of a kernel function is the linear kernel, defined as $K_\text{lin}(\mathbf{x},\mathbf{y})=\mathbf{x}^T\cdot\mathbf{y}$, where the dot product is taken between two descriptor vectors, $\mathbf{x}$ and $\mathbf{y}$.\cite{bishop2006pattern} We will use a linear kernel function throughout this work. However, there are numerous other forms of kernel functions that are commonly used. To describe each local atomic environment, we employ the Smooth Overlap of Atomic Positions (SOAP) descriptors developed by Bart{\'o}k {\em et al.}\cite{bartok2013representing}, together with a Behler--Parrinello cutoff function to ensure locality and linear scaling with the number of atoms in the system.\cite{behler2007generalized} For the real spherical harmonics, we used the implementation of  \texttt{sphericat}.\cite{sphericart}

During training, overall charge neutrality can be enforced through the constraint
\begin{equation}\label{eq:const}
\sum_i q_i = 0 \;\Rightarrow\; \sum_i \sum_{I_B} \omega_{I_B} K(X_i, X_{I_B}) = 0.
\end{equation}
This constraint is applied to the training configurations and is therefore satisfied once the optimisation has converged. Charge neutrality for unseen structures is not guaranteed unless the constraint is explicitly enforced at prediction time or built into the model by construction.

Dipole moments $d^\alpha_i$ are modelled as vector quantities associated with each ion $i$. The total polarisation of the system can be expressed as the sum over all individual dipole contributions:
\begin{equation}
    P^\alpha = \sum_i d^\alpha_i.
\end{equation}
Analogous to the monopole case, the BECs quantify how the polarisation responds to atomic displacements. They are defined as the derivative of the polarisation with respect to the atomic positions, as in Eq.~\eqref{eq:Zdie}. In our experience, a practical limitation of a linear kernel is that they cannot represent constant offsets well. As a result, the diagonal components of the BEC tensor cannot be learned directly within the dipole kernel model. To account for this, constant shifts are introduced and fitted alongside the remaining model parameters. These constant terms are mathematically equivalent to constant monopole contributions and ensure that the diagonal components of the BECs are described accurately.

In the machine learning framework, dipoles are predicted using symmetry-adapted descriptors that transform as vectors under rotations, known as $\lambda$-SOAP developed by Grisafi {\em et al.}\cite{grisafi2018symmetry} The dipole is modelled as
\begin{equation}
    d^\alpha_i = \sum_{I_B} \omega_{I_B} K^\alpha(X^\alpha_i,X_{I_B}).
\end{equation}
This dipole-only formulation constitutes the methodology developed and employed in our previous work (Ref.~\onlinecite{schmiedmayer2024derivative}). In the results section, we treat this model as the benchmark against which the scalar monopole model is evaluated.

The monopole and dipole models can be combined to improve the accuracy of the machine learning framework for predicting BECs. The combined expression for the BEC tensor is then given by:
\begin{equation}
    \begin{split}
    Z^{\alpha\beta}_i = \sum_{I_B} \omega_{I_B}\left(K(X_i,X_{I_B})\delta^{\alpha\beta} + \sum_j \frac{\dd K(X_j,X_{I_B})}{\dd r^\beta_i}\left(r_j^\alpha-r_i^\alpha\right)\right)\\ +\sum_{J_B} \omega_{J_B}\sum_j\frac{\dd K^\alpha(X^\alpha_j,X_{J_B})}{\dd r^\beta_i}.
    \end{split}
\end{equation}

\subsection{Neural network}
As an example of BEC fitting using equivariant graph neural networks, the monopole model defined in Eq.~\eqref{eq:Zsol} is implemented in the MACE architecture.\cite{batatia2022mace} The scalar output of each atom graph is taken as the monopole charge $q_i$. The full derivatives of $q_j$ with respect to the atomic positions $r^\beta_i$ are evaluated by automatic differentiation.\cite{bucker2006bibliography} The BECs are then constructed according to
\begin{equation}
  Z_{i,\alpha\beta}^{*,\text{MACE}}(r^\alpha_i) = q_i\delta^{\alpha\beta} + \sum_{j} \frac{\dd q_j}{\dd r^\beta_i}(d_{ij}^\alpha).
\end{equation}
The sum is taking into account all atoms, and $d_{ij}$ defines the distance between atom $i$ and $j$ within the minimum image convention. The models are then trained on the mean square error of the predicted BECs. To ensure charge neutrality explicitly, each monopole charge is shifted by the mean of all charges before evaluating the derivatives.

All results in this work were obtained from MACE models featuring two layers with 32 channels, passing only invariant messages (32$\times$0e). No increase in performance was to be gained by including equivariant messages (L$_{\mathrm{max}}>0$), as one would expect when restricting the BEC model to monopole contributions only. Models were trained for 250 epochs, with a validation set of at least 5 configurations used to select the best performing model. The only system-specific hyperparameter was the cut-off radius, set to \SI{4}{\angstrom} for \ce{H2O}, \SI{5}{\angstrom} for \ce{ZrO2} and \ce{MAPbI3}, and \SI{6}{\angstrom} for \ce{NaCl}.   

\section{Results and Discussion}

\subsection{Model comparison}

To compare the effectiveness of the three models, we employed four BEC datasets of bulk materials. The three models are: a monopole model ($q$), where the BEC are represented by environment-dependent atomic charges [Eq.~\eqref{eq:Zsol}]; a dipole model ($p$), where the BEC is captured through learned atomic dipole moments [Eq.~\eqref{eq:Zdie}]; and a combined monopole--dipole model ($q+p$), which includes both contributions [Eq.~\eqref{eq:Z}]. In addition, the monopole model is also evaluated within the MACE framework, which represents a more expressive machine-learning architecture than the kernel-based approach. The first two datasets correspond to liquid water at room temperature and to the orthorhombic, tetragonal, and cubic phases of \ce{MAPbI3}, which were both prepared in Ref.~\onlinecite{schmiedmayer2024derivative}. The third dataset, consisting of liquid \ce{NaCl} at temperatures between \SIrange[range-phrase = { and }]{1100}{1400}{\kelvin} and was taken from Ref.~\onlinecite{faller2024density}. Only the atomic configurations were adopted from this reference; the BECs were computed in the present work. The fourth and final dataset, comprising \ce{ZrO2}, was generated for this work. The BECs were computed using density functional perturbation theory (DFPT) \cite{wu2005systematic,gajdovs2006linear,perez2015vibrational} by evaluating the static ion-clamped dielectric matrix, following the approaches of Baroni and Resta \cite{baroni1986ab} and Gajdoš \textit{et al.} \cite{gajdovs2006linear} for the projector-augmented wave (PAW) method as implemented in \textsc{VASP}. \cite{kresse1996efficiency,kresse1996efficient,kresse1999ultrasoft} In total, \num{119} structures covering the monoclinic, tetragonal, and cubic phases of \ce{ZrO2} and temperatures between \SIrange[range-phrase = { and }]{500}{1600}{\kelvin} are included in this dataset.

The liquid water dataset consists of \num{100} configurations, each containing \num{64} molecules, resulting in a total of \num{57600} fit equations. The liquid \ce{NaCl} dataset contains \num{134} configurations with \num{128} atoms, resulting in a total of \num{51456} fit equations. The \ce{MAPbI3} dataset consists of \num{300} structures with \num{96} atoms, resulting in a total of \num{86400} fit equations. 

\begin{figure*}
    \centering
    \includegraphics{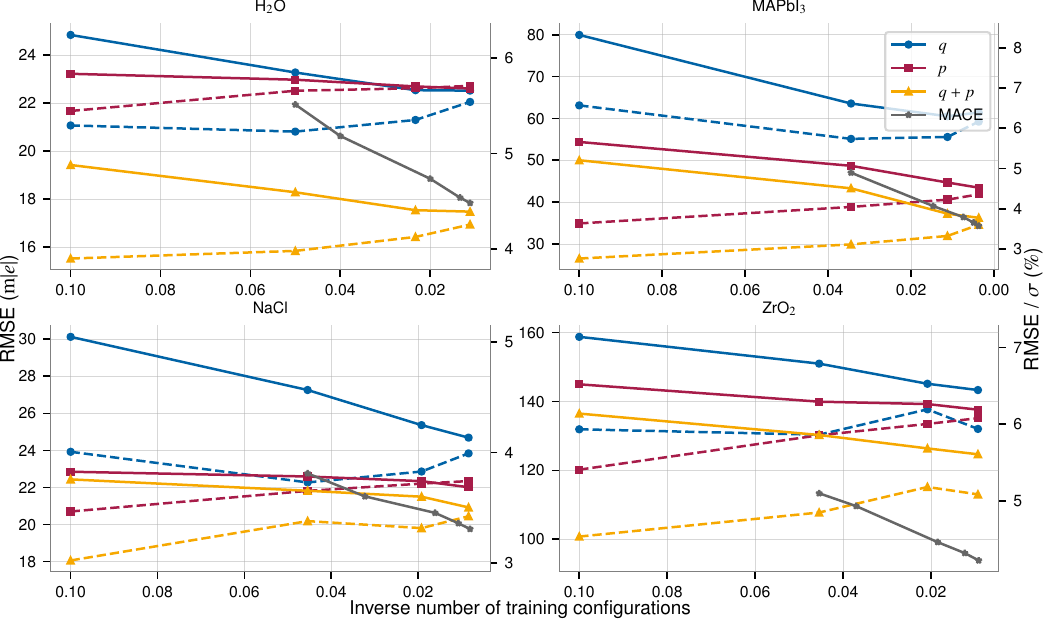}
    \caption{Root mean square error (RMSE), left axes and normalised root mean square error (RMSE divided by the standard deviation of the values in the reference DFT data set, right axes) on a test set (\SI{10}{\percent} of the total data) as a function of the number of training configurations for three models: $q$--monopoles (circle), $p$--dipoles (square), $q+p$--combined monopoles and dipoles (triangle), and MACE (star). Results are shown for data sets on liquid water at room temperature (\ce{H2O}), the orthorhombic, tetragonal, and cubic phases of \ce{MAPbI3}, liquid \ce{NaCl} between \SIrange[range-phrase = { and }]{1100}{1400}{\kelvin}, and solid \ce{ZrO2}. Broken lines indicate training set errors, while the solid lines are the test set errors (for MACE only test set errors are shown, while the number of configurations refers to the combined training and validation set).}
    \label{fig:LearningCurve}
\end{figure*}

To assess the performance of the different models --- monopoles, dipoles, and their combined training --- we generated learning curves for all four datasets. A test subset comprising \SI{10}{\percent} of each dataset was held out from training and used exclusively to assess model performance. At each data point, a hyperparameter optimisation was performed for the kernel model to minimise the test set error. The resulting learning curves are shown in Fig.~\ref{fig:LearningCurve}.

Firstly, it should be noted that the test set errors decrease systematically as the size of the training set increases, whereas the training set errors usually increase slightly when the number of training structures increases. They both converge to very similar values, implying that we have reached the expressiveness limit of the simple linear regression models used in this study. However, the root mean square errors are very good, usually around or even below \SI{5}{\percent} of the standard deviation of the training data. We consider the residual error to be acceptable, and as demonstrated below, there are no significant differences between the different models for the infrared spectra.

Overall, it can be observed that the combined training of the monopole and dipole models yields the lowest test set error across all systems among the SOAP based, simple linear-regression models. This is expected, as it represents the most comprehensive and flexible model, incorporating both the monopole and dipole contributions to the multipole expansion of $\mathbf{Z}^*$. While the dipole model (red lines) generally performs better than the monopole model, some nuances are worth noting. The dipole curves ($p$) in Fig.~\ref{fig:LearningCurve} represent the performance of the methodology from Ref.~\onlinecite{schmiedmayer2024derivative}. While this tensorial approach generally yields a lower error than the scalar monopole model for small datasets, the scalar model approaches comparable accuracy with sufficient data. The learning curves of the monopole model exhibit a steeper slope, whereas those of the dipole model start at a lower error but hardly improve with the training set size. Comparing the convergence of training and test set errors with an increasing number of training data suggests that the monopole model reaches a slightly higher asymptotic error for most systems. An exception is \ce{H2O}, where the monopole model performs marginally better for larger datasets, though the difference remains small. Overall, the dipole-only model is extremely data efficient. However, in its current linear regression form, it is not flexible enough to produce errors much smaller than \SI{5}{\percent}.
The combined model has a similar data efficiency and reduces the overall errors somewhat.

The test set error of the scalar MACE model is lower than that of the kernel method for all systems, with the exception of water. However, the slope of the learning curves suggests that only a modest increase in the number of training configurations would be sufficient for MACE to achieve a lower error, for all datasets. This indicates that a more expressive machine-learning framework can surpass more elaborate models --- such as the combined monopole--dipole approach ($q+p$) --- implemented within a less flexible machine-learning architecture. The greater accuracy of the MACE model is probably due to the inclusion of higher-order many-body terms (the SOAP descriptors used here only include two- and three-body terms) and the longer range associated with the single message-passing layer. 

In conclusion, for simple linear regression models, the combined model demonstrates the best overall performance, followed by the dipole-only model. The monopole model approaches similar accuracy but requires more training data to achieve comparable test set errors. Nevertheless, the monopole model remains attractive due to its simplicity and ease of integration in scalar machine-learning frameworks. When embedded in a more expressive architecture such as MACE, the monopole model is able to outperform all linear regression models considered in this work, but not by a significant margin. Likely larger training sets would tip the balance even more towards the equivariant message-passing model.

In addition to the bulk systems, we evaluated the model performance on a water dimer (\ce{2(H2O)}), using Eq.~\eqref{eq:Zmol} in the kernel-based model to learn $\mathbf{Z}^*$. For this system, all models achieved very low training errors in the range of \SIrange{6.0}{8.1}{\milli|\elementarycharge|}, with the dipole model showing slightly better performance than the monopole model.

\subsection{Point charge analysis}

\begin{figure}
    \centering
    \includegraphics{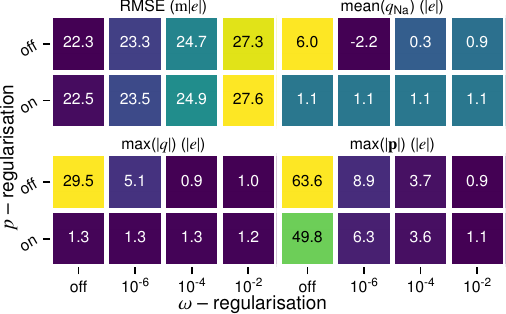}
    \caption{Root mean square error (RMSE), mean atomic charge of \ce{Na} ($\mathrm{mean}(q_{\ce{Na}})$), maximum absolute atomic charge ($\mathrm{max}(|q|)$), and maximum absolute atomic polarisation over all Cartesian components ($\mathrm{max}(|\mathbf{p}|)$) as a function of different regularisation schemes for the liquid \ce{NaCl} dataset for the combined model ($q+p$). Predicted values are reported alongside colour coding of that value. Two types of regularisation are shown: Tikhonov regularisation (x-axes $\omega$-regularisation), which penalises large fitting parameters $\omega$, and a charge regularisation that constrains the atomic charges $q$ to reproduce $\mathrm{trace}(\mathbf{Z}^*)/3$ (y-axes $p$-regularisation).}
    \label{fig:Reg_Scan_p}
\end{figure}

The effective dipole moment of a water molecule in the gas phase in the principal axis energy representation has been measured from Stark effects to be \SI{-0.386}{\angstrom|\elementarycharge|}.\cite{shepard1973dipole} Taking the structure of water to be rigid at its experimental geometry, $r_{\ce{OH}}=\SI{0.9572}{\angstrom}$ and $a_{\ce{HOH}}=\SI{104.523}{\degree}$,\cite{benedict1956rotation} and assuming $\mathrm{C}_\mathrm{2v}$ symmetry, the effective atomic charges are computed to be $q(\ce{O}) = -\SI{0.66}{|\elementarycharge|}$ and $q(\ce{H}) = +\SI{0.33}{|\elementarycharge|}$.\cite{martin2005charge} The predicted average monopoles of the oxygen atoms in liquid water are $q(\ce{O}) = \SI{-0.37}{|\elementarycharge|}$ using the monopole SOAP model, and  $q(\ce{O}) = \SI{-0.54}{|\elementarycharge|}$ for MACE. On the other hand, for the water dimer, the dipole-only based $\lambda$-SOAP framework yields an average oxygen charge of $q(\ce{O}) = \SI{-0.60}{\elementarycharge}$ --- extracted from the fitted diagonal elements of the BEC tensor --- which is in much closer agreement with the experimentally deduced value. Generally, the predicted atomic charges deviate by up to a factor of two from the experimentally deduced values. This discrepancy is not unexpected. While the nuclei can be treated as positive point charges, the electrons exist as a distributed cloud over the entire molecule. A point charge model is a simplification that replaces the continuous charge distribution with a few discrete charges. Because these partial charges are parameters of the model --- chosen to reproduce the correct BEC --- they are not themselves fundamental, physical quantities. They are an artefact of the simplified model, so it is expected that their specific values are not physically meaningful or interpretable. It is also important to note that hyperparameter choices can strongly influence the fitted charges (see Fig.~\ref{fig:Reg_Scan_p}). Moreover, different theoretical levels and charge localisation schemes can also yield widely varying charge assignments even for a simple \ce{H2O} molecule.\cite{martin2005charge} Comparable charge distributions can nevertheless be achieved when placing point charges at positions other than the atomic sites.\cite{izadi2014building} 

To strengthen this conclusion, Fig.~\ref{fig:Reg_Scan_p} shows the sensitivity of the atomic charges and dipoles to the choice of the regularisation for the liquid \ce{NaCl} dataset, together with the corresponding RMSE. We show the RMSE, the mean atomic charge of \ce{Na} ($\mathrm{mean}(q_{\ce{Na}})$), the maximum absolute atomic charge ($\mathrm{max}(|q|)$), and the maximum absolute atomic polarisation ($\mathrm{max}(|\mathbf{p}|)$) in four sub panels. Two qualitatively different regularisation strategies are compared.

The first strategy is a standard Tikhonov regularisation, which penalises large fitting parameters $\omega$ (variations within a subpanel along x-axes ``$\omega$-regularisation'' in Fig.~\ref{fig:Reg_Scan_p}). Varying the strength of this regularisation has a pronounced effect on the magnitude of both the learned atomic charges and dipoles: stronger regularisation systematically suppresses their absolute values, while weaker regularisation allows them to grow substantially. This behaviour highlights an important conceptual point that we already raised above: the individual atomic charges and dipoles obtained from the model are not uniquely defined. In particular, even for a highly ionic system such as \ce{NaCl}, where one would intuitively expect an average sodium charge close to $+$\SI{1}{|\elementarycharge|}, the learned charges can deviate strongly from this value depending on the regularisation strength. This arbitrariness is expected, as the model is trained only on BEC, {\em i.e.}, on the change of the total polarisation with respect to atomic displacements, rather than on the absolute magnitude of the polarisation itself or the actual physical charge distribution. With access only to this differential information, there is no unique way to fix the absolute scale of the atomic charges or dipoles. A more stringent Tikhonov regularisation systematically increases the RMSE, but not substantially so.

The second regularisation scheme directly constrains the atomic charges $q$ to reproduce $\mathrm{trace}(\mathbf{Z}^*)/3$ (variations within a subpanel along y-axes ``$p$-regularisation'' in Fig.~\ref{fig:Reg_Scan_p}). This additional constraint effectively fixes the otherwise undetermined monopole contribution and yields sodium charges close to the expected ionic value, while increasing the RMSE only marginally. This demonstrates that a regularisation can recover chemically intuitive charge magnitudes without compromising the quality of the BEC prediction. Importantly, no analogous constraint for the dipole moments $\mathbf{p}$ was implemented, whose absolute magnitude therefore remains inherently ambiguous within the present framework. Overall, these results underscore that while the monopole--dipole decomposition is a useful and flexible representation of the BEC, the individual monopole and dipole contributions are not related to true spatial charge density distribution changes as one would observe in first principles calculations, but rather as model-dependent constructs whose values depend sensitively on the model's hyperparameters. They only fulfill the requirement to describe the BECs as accurately as possible. 

Recent studies have suggested that the BECs can also be deduced from long-range interactions.\cite{cheng2025latent} This is certainly true, as the long-range interactions are fully determined by dipole--dipole interactions (see, for example, Ref.~\onlinecite{ghosez1997dynamical}),
\begin{equation}
    \frac{ (\mathbf{Z} \delta \mathbf{r}) (\mathbf{Z} \delta \mathbf{r}')} {\epsilon |\mathbf{r} - \mathbf{r}'|^3 }. 
\end{equation}
Consequently, such interactions contain similar information to that obtained from explicitly calculated BECs, albeit screened by the dielectric constant, and only if the training structures contain hundreds of atoms and are sufficiently large. However, we strongly doubt this strategy would yield more ``correct'', physically interpretable local charges, since the electrostatic long-range interactions are always combined with a short-range force field. Therefore, the charges obtained by fitting energies are unlikely to represent true charge-density rearrangements. As in the present study, they will only yield the correct long-range mesoscopic charge rearrangement ($\mathbf{Z}\delta\mathbf{r}$) and any unphysical charge ``decomposition'' will be fully compensated in the short and medium range by the short-range force field. In short, regression is not a ``magical'' method to uncover underlying physics; it rather attempts to fit the desired quantities as well as possible.

\subsection{Infrared spectra}
To measure the IR spectrum, several experimental techniques can be employed. The experimental data referenced in this work were obtained using attenuated total reflectance (ATR) spectroscopy for water\cite{bertie1996infrared} and transmission spectroscopy for \ce{MAPbI3}.\cite{schuck2018infrared} In both cases, the frequency-dependent absorption coefficient $\alpha(\omega)$ is measured. The absorption coefficient is defined as\cite{dresselhaus2018solid,schmiedmayer2024derivative}
\begin{equation}
\alpha(\omega) \approx \frac{\beta}{3V\epsilon_0c} \int_0^\infty \langle \dot{\mathbf{P}}(0) \cdot \dot{\mathbf{P}}(t) \rangle \cos(\omega t)\mathrm{d}t,
\end{equation}
where $V$ is the system volume, $\epsilon_0$ the vacuum permittivity, $c$ the speed of light in vacuum, and $\beta = 1/(k_\mathrm{B}T)$ the inverse thermal energy, with $k_\mathrm{B}$ being the Boltzmann constant and $T$ the temperature.

The time derivative of the total polarisation, $\dot{\mathbf{P}}$, is obtained from
\begin{equation}
\frac{\mathrm{d} P^\alpha}{\mathrm{d}t} = \sum_{i,\beta} \frac{\partial P^\alpha}{\partial x_i^\beta} \frac{\mathrm{d} x_i^\beta}{\mathrm{d}t} = \sum_{i,\beta} Z_i^{*\alpha\beta} v_i^\beta,
\end{equation}
where $Z_i^{*\alpha\beta}$ are the BEC tensors and $\mathbf{v}_i$ are the atomic velocities, computed as numerical gradients from the molecular dynamics (MD) trajectories reported in Ref.~\onlinecite{schmiedmayer2024derivative}.

\begin{figure}
    \centering
    \includegraphics{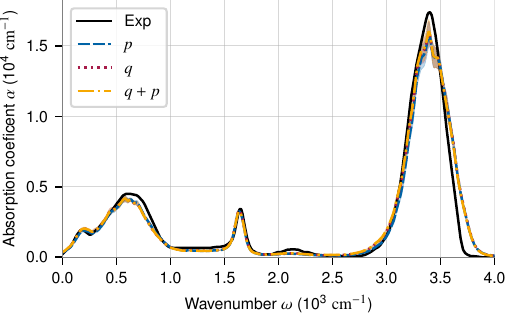}
    \caption{Experimental and computational IR spectra of liquid water. The computational spectra were obtained using three different models: $q$--monopoles (dashed), $p$--dipoles (dotted), $q+p$--combined monopoles and dipoles (dash-dot). Experimental reference data are taken from Ref.~\onlinecite{bertie1996infrared}. Statistical uncertainties of the simulations are indicated by the shaded regions surrounding the calculated spectra, corresponding to the \SI{95}{\percent} confidence interval, {\em i.e.}, $\pm 2 \sigma$, where $\sigma$ is the standard error of the sample mean.}
    \label{fig:specH2O}
\end{figure}

The computed IR spectrum of liquid \ce{H2O} at room temperature is shown alongside experimental data in Fig.~\ref{fig:specH2O}. The spectrum was obtained by averaging \num{20} independently computed IR spectra. The underlying molecular dynamics trajectories were generated using the RPBE+D3 functional.\cite{hammer1999improved,grimme2010consistent} Each individual spectrum was calculated from a microcanonical (NVE) MD trajectory. For each run, the simulation was initialised from an uncorrelated configuration, with initial velocities sampled from a canonical ensemble using a Langevin thermostat to ensure equilibration at room temperature. After equilibration, each trajectory was propagated for \num{100000} MD steps with a time step of \SI{0.25}{\femto\second}. The use of multiple trajectories starting from uncorrelated initial conditions improves the statistical reliability of the resulting spectrum. Prior to performing the Fourier transform, a Gaussian window function was applied to the dipole--dipole autocorrelation function instead of Lorentzian broadening. This choice yields sharper spectral features in the computed IR spectra.

As shown in Fig.~\ref{fig:specH2O}, the present methodology enables the computation of the IR spectrum of liquid water with remarkable agreement with experimental data for all three modelling approaches. Three important conclusions can be drawn from this result. First, all three approaches --- monopole, dipole, and the combined monopole--dipole model --- produce nearly identical IR spectra. This indicates that the some what larger instantaneous errors present in the monopole model are effectively averaged out through thermodynamic averaging of the dipole--dipole autocorrelation function. Second, the close agreement in spectral intensities with experiment reflects the accurate description of the BECs. Since the models are explicitly trained on BECs, this level of agreement is expected. Third, the RPBE+D3 functional provides an excellent description of the dynamics of liquid water, accurately capturing both the high-frequency \ce{O-H} stretching modes and the intermediate-frequency bending modes. We note that hybrid functionals lead to an increase of the \ce{O-H} stretch frequency whereas quantum statistics has an opposite effect, so that the good agreement is to some extent fortuitous.\cite{schmiedmayer2024derivative} Regardless of this, low-frequency spectral features associated with intermolecular motions are well reproduced, highlighting the quality of the underlying DFT functional and molecular dynamics simulations.

\begin{figure*}
    \centering
    \includegraphics{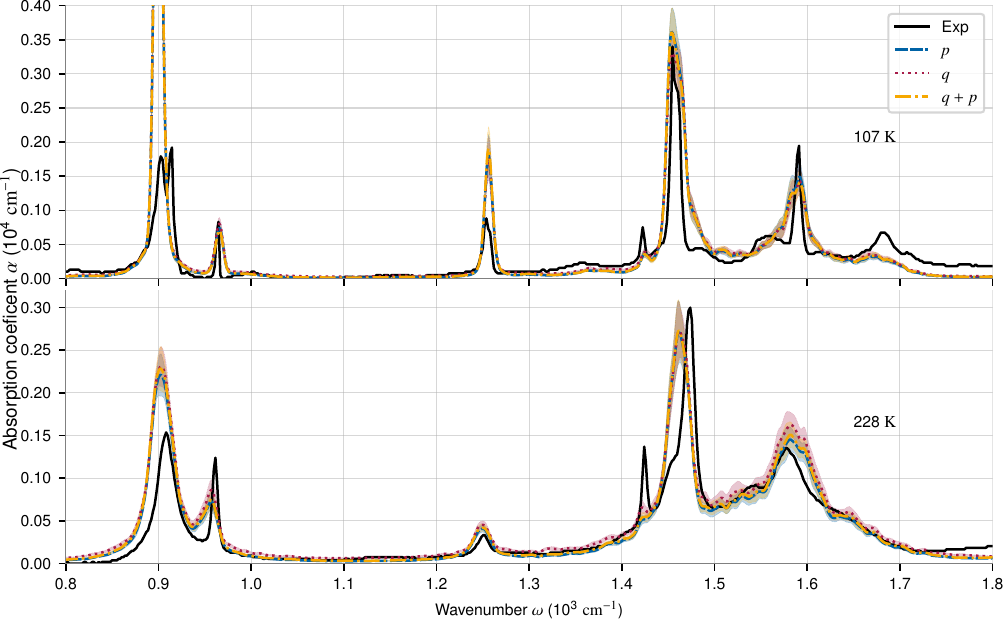}
    \caption{Experimental and computational IR spectra for the orthorhombic and tetragonal phases of \ce{MAPbI3}. The computational spectra were obtained using three different models: $q$--monopoles (dashed), $p$--dipoles (dotted), and the combined $q+p$ monopole--dipole model (dash--dot). The vibrational frequencies of the computed spectra have been uniformly redshifted by \SI{1.5}{\percent} to align with the experimental data. Experimental reference spectra are taken from Ref.~\onlinecite{schuck2018infrared} and are shown in arbitrary units, as only relative intensities are reported. Statistical uncertainties of the simulations are indicated by the shaded regions surrounding the calculated spectra, corresponding to the \SI{95}{\percent} confidence interval, {\em i.e.}, $\pm 2\sigma$, where $\sigma$ is the standard error of the sample mean.}
    \label{fig:specMAPBI}
\end{figure*}

In Fig.~\ref{fig:specMAPBI}, we present the computed IR spectra of \ce{MAPbI3} for the orthorhombic phase at \SI{107}{\kelvin} and the tetragonal phase at \SI{228}{\kelvin}, together with the corresponding experimental spectra for comparison. The spectra for both phases were calculated using a $4 \times 4 \times 4$ supercell in order to improve statistical sampling and to allow for orientational disorder and rearrangement of the methylammonium molecules. The machine-learning force field (MLFF) was trained on SCAN-based reference data.\cite{sun2015strongly} The initial configurations for the individual MD trajectories were sampled from an isothermal--isobaric ensemble. Apart from using starting configurations with different cell vectors, the procedure for computing the IR spectra closely follows that described above for liquid water.

Consistent with the results obtained for liquid water, all three approaches --- monopole, dipole, and the combined monopole--dipole model --- yield nearly identical IR spectra for \ce{MAPbI3}. Crucially, the spectra generated by the new scalar monopole model are virtually indistinguishable from those generated by the tensorial dipole model used in our previous study (Ref.~\onlinecite{schmiedmayer2024derivative}). This confirms that the scalar approximation captures the necessary physics for spectroscopic predictions despite the simplified descriptors. Overall, the computed spectra show very good agreement with the experimental reference; however, some discrepancies remain, most notably in the relative intensities of individual peaks. In particular, the modes around \SI{900}{\per\centi\meter} appear with systematically higher intensity in the simulations.

Such intensity deviations may partly originate from residual errors in the predicted BECs, which directly enter the calculation of IR intensities. In addition, inaccuracies in the MLFF can affect both peak positions and intensities through subtle changes in the underlying vibrational dynamics. Within the present framework, these contributions are difficult to disentangle. It should also be noted that the experimental reference spectra were obtained from a single crystal,\cite{schuck2018infrared} and may therefore be influenced by surface effects and crystal orientation, which can further affect the measured intensities. A more detailed analysis of the IR spectra of \ce{MAPbI3}, including a comparison with DFPT results, is provided in a previous publication in Ref.~\onlinecite{schmiedmayer2024derivative}.

\section{Conclusion}
In this study, we have revisited the dipole-based framework introduced in our previous work (Ref.~\onlinecite{schmiedmayer2024derivative}) and demonstrated that similar results can be obtained using a simpler, purely scalar machine-learning formulation. Expressing the BEC tensor via a monopole--dipole response decomposition allowed us to show that invariant descriptors alone are sufficient to reproduce the correct polarisation response and finite-temperature infrared spectra across a range of complex materials.

At the level of individual BECs, the monopole-based model exhibits larger prediction errors than the dipole and combined monopole-dipole approaches. Nevertheless, it remains a sufficiently reliable approximation in practice. When employed in molecular dynamics simulations, these errors are effectively averaged out in the IR spectrum, as evidenced by the virtually indistinguishable IR spectra obtained with all three models. It is important to emphasise that the monopole charges associated with individual atoms are seemingly not physically meaningful but rather model-dependent parameters introduced to reproduce the correct polarisation response. Their absolute atom-centred values, therefore, carry no direct physical meaning, except that the sum of all their changes reproduces the target BECs.

A direct comparison between the monopole and dipole models is not entirely equitable, as the dipole-based approach relies on constant monopole contributions. Because linear kernel models struggle with representing constant terms, the diagonal contribution to the BEC tensor must be removed during preprocessing. These implicit monopole terms are essential for achieving low prediction errors and further complicate a strict one-to-one comparison. Disregarding the fixed monopole terms, the dipole model, as formulated in Ref.~\onlinecite{schmiedmayer2024derivative}, is more ``interpretable'' in that it learns atomic dipoles whose sum defines the total polarisation of the system but for a constant, undefined off-set vector. The BECs then strictly follow as derivatives with respect to atomic displacements. This guarantees the existence of a global anti-derivative, {\em i.e.} the system's polarisation. By contrast, the monopole model does not define a unique global polarisation because each BEC is constructed with respect to an individually chosen local centre.

Despite these limitations, the monopole-based formulation offers clear practical advantages. Its conceptual simplicity, reliance on scalar descriptors, and absence of explicit tensorial equivariance make it straightforward to integrate into existing scalar machine-learning frameworks. When combined with more expressive architectures such as MACE, the monopole model can even outperform kernel-based implementations including dipoles. Overall, these results establish the monopole approach as a robust, scalable, and easily deployable alternative for large-scale simulations, particularly in contexts where computational efficiency and compatibility with existing ML infrastructures are key considerations. Finally, simple monopole models are widely used in semi-empirical quantum chemistry, as well as for embedding high-level quantum chemistry into simpler electrostatic models. Making the monopole charges fully trainable could greatly enhance the accuracy of electrostatic embedding. Monopole models are closer to the core of quantum chemistry than the dipole-based approach, which physicists often use to describe long-range interactions in covalent solids. Monopole models' ability to describe BECs with a level of accuracy similar to dipole models has significant implications for our understanding of materials' electrostatics. However, as already emphasized, the atom-centered monopoles do not appear to be particularly well defined and are highly dependent on regularisation parameters. Therefore, there is no specific physics attached to them; they are merely model parameters that describe long-range electrostatics. Ideally, we should be able to link or constrain the atom-centred monopole and dipole charges to actual local changes in the electronic charge distribution. This would allow us to build intermediate-range electrostatic models and make the results more easily interpretable. However, in the current era of data-driven science, it is uncertain whether the additional complexity of fitting real (first principles)-derived charge-density changes will ever be competitive with a more data-driven approach based on a small number of labels, such as energies, forces, and BECs.


\section{Acknowledgements}
This research was funded in whole by the Austrian Science Fund (FWF) 10.55776/F8100 and 10.55776/COE5. For open access purposes, the author has applied a CC BY public copyright license to any author accepted manuscript version arising from this submission. The computational results presented have been achieved in part using Austrian Scientific Computing (ASC) resources.



\bibliography{ref.bib}
\end{document}